\documentclass[11pt]{article}
\usepackage{moriond,epsfig}

\def\Journal#1#2#3#4{{#1} {\bf #2}, #3 (#4)}

\def\NPB{{\em Nucl. Phys.} B}
\def\PLB{{\em Phys. Lett.}  B}
\def\PRL{\em Phys. Rev. Lett.}

\def\be{\begin{equation}}
\def\ee{\end{equation}}
\def\bea{\begin{eqnarray}}
\def\eea{\end{eqnarray}}

\def\to{\rightarrow}
\def\Kl{K^0_L}
\def\Ks{K^0_S}
\def\ppee{\pi^0\pi^0e^+e^-}
\def\mmee{\mu^+\mu^-e^+e^-}
\def\eeee{e^+e^-e^+e^-}
\def\mmg{\mu^+\mu^-\gamma}
\def\pen{\pi^\pm e^\mp\nu}
\def\row{\rho_{\mathrm CKM}}
\def\ate{\eta_{\mathrm CKM}}
\def\CP{\hbox{\rm CP}}
\def\CPT{\hbox{\rm CPT}}
\def\Pt2{{\mathrm P}_\perp^2}
\def\vect{\overrightarrow}
\def \pma#1#2 {\mbox{\raisebox{-0.6ex}
           {$\stackrel{\scriptstyle \;+\; #1}{\scriptstyle \;-\; #2}$}}}

\begin{document}
\vspace*{4cm}
\title{RECENT RESULTS FROM KTEV}

\author{ L. BELLANTONI for KTEV }

\address{Fermi National Accelerator Laboratory,\\
Batavia IL, U.S.A.}

\maketitle\abstracts{
Recent results are presented for (1) the charge asymmetry in
semielectronic kaon decay; (2) the charge radius of the neutral kaon;
(3) the decay $\Kl\to\ppee$; (4) constraints on $\row$ from kaon decays;
(5) lepton flavor violation.  A few words about future
kaon physics work at Fermilab are included.}

\section{Charge Asymmetry}

The charge asymmetry in the decay $\Kl\to\pen$, defined by
\begin{equation}
\delta_L = \frac{N(\Kl\to\pi^-e^+\nu)-N(\Kl\to\pi^+e^-\overline\nu)}
                {N(\Kl\to\pi^-e^+\nu)+N(\Kl\to\pi^+e^-\overline\nu)}
\end{equation}
equals $2\Re(\varepsilon-Y-X_-)$, where $\varepsilon$ is the familiar
indirect $\CP$ violation parameter, $Y$ parameterizes $\CPT$ violation
in $\Delta S = \Delta Q$ amplitudes and $X_-$ parameterizes $\CPT$
violation in $\Delta S = -\Delta Q$ amplitudes \cite{daphne}.  A
comparison of the real part of $\varepsilon$ as measured in
$K\to\pi\pi$ decays with  $\delta_L$ can reveal new, $\CPT$ violating
processes.

The PDG 2000 average for $e^\pm$ and $\mu^\pm$ is
$\delta_L = (3.27 \pm 0.12) \times 10^{-3}$; the best $e^\pm$
result \cite{vera} is $(3.41 \pm 0.18) \times 10^{-3}$, obtained
from a dataset of $34 \times 10^6$ events.  Here we present a
result based on about $300 \times 10^6$ events.

Building a detector with the same efficiencies for both charge
combinations at the required level is approximately impossible.
Instead, we begin by defining subsamples with nearly cancelling
efficiencies.  So for example we compare the number of $e^+$ events
measured when the spectrometer magnet was set to positive polarity to
the number of $e^-$ events with negative magnet polarity.  Then,
\begin{equation}
R = \frac{N(\Kl\to\pi^-e^+\nu;+mag)}
         {N(\Kl\to\pi^+e^-\overline\nu;-mag)}
  = \frac{Br(\to e^+) \cdot Flux(+mag) \cdot \epsilon(e^+;+mag)}
         {Br(\to e^-) \cdot Flux(-mag) \cdot \epsilon(e^-;-mag)},
\end{equation}
and the ratios of dataset size $\frac{Flux(+mag)}{Flux(-mag)}$ and
detection efficiency
$\frac{\epsilon(e^+;+mag)}{\epsilon(e^-;-mag)}$ nearly
cancel.  Since the KTeV apparatus contains two beams, there are four
values of $R$; we take the fourth root of their product.

Clearly, the viability of the method lies in our ability to allow for
small variations from perfect cancellation.  Table~\ref{tab:corrs} lists
the corrections we have made; a detailed paper is being
written \cite{hogan}.  Our
preliminary result on the data taken in 1997 is
\begin{equation}
\delta_L = (3.320 \pm0.058_{STAT} \pm0.046_{SYS}) \times 10^{-3}.
\end{equation}
With this result, the new world average is
$(3.320 \pm0.063) \times 10^{-3}$, which differs from
$(2\Re(\eta_{+-}) + \Re(\eta_{00}))/3$ by only
$(-2\pm35) \times 10^{-6}$, consistent with $\CPT$ conservation.

\begin{table}[t]
\caption{Summary of corrections for systematic effects in
$\delta_L$, in units of $10^{-8}$, with sources for the corrections.
\label{tab:corrs}}
\vspace{0.4cm}
\begin{center}
\begin{tabular}{|l|c|r|}
\hline
$\pi^+\pi^-$ different in calorimeter       & Data       &$-156\pm10$\\
$\pi^+\pi^-$ loss in trigger scintillator   & Data       &  $54\pm10$\\
$\pi^+\pi^-$ loss in spectrometer           & Simulation &   $3\pm 3$\\
$\pi^+\pi^-$ punchthrough differences       & Data       &  $34\pm40$\\
$e^+ e^-$ different in calorimeter          & Data       & $-19\pm18$\\
$\delta$ ray production differences         & Simulation &  $-8\pm 4$\\
$e^+$ annihilation in spectrometer          & Simulation &  $11\pm 1$\\
Backgrounds ($K_{\pi3}, K_{\mu3}, \Lambda$) & Data,Simulation &
                                                             $1\pm 1$\\
Target/adsorber interference                & Simulation &  $-2\pm 1$\\
Collimator, regenerator scatters of $\Kl$   & Data       &  $-1\pm 1$\\
Spectrometer polarity mismatch              & Data       &  $-3\pm 2$\\
\hline
Total                                       &           &  $-97\pm46$\\
\hline
\end{tabular}
\end{center}
\end{table}

\section{Charge Radius}

There are three contributions to the process  $\Kl\to\pi^+\pi^-e^+e^-$.
The first is a direct emission amplitude (DE) where the $\Kl\pi\pi$
vertex emits a photon; the second is the bremsstrahlung amplitude (BR);
and the third is the charge radius (CR) amplitude.  This third term
corresponds to a $\Kl\to\Ks\gamma^*$ transition immediately prior to
a $\Ks\pi\pi$ vertex.  The photon then materializes as an $e^+e^-$
pair.  Each of these amplitudes are multiplied by a coupling factor,
and the factor multiplying the CR term is proportional to the charge
radius, $
-\left\langle \Sigma q_i(\vect r - \vect R)^2 \right\rangle m_K^2 /3$.
We use our clean sample of 1811 events \cite{brad} to fit the charge
radius amplitude's coupling constant using the data's phase space
distribution while holding the coupling constants for the other
amplitudes, the indirect $\CP$ violation parameters $\eta_{+-}$ and
$\Phi_{+-}$ and the DE form factor fixed.  Variations in these
parameters are used to assign systematic uncertainties, as
is the finite size of the Monte Carlo sample, potential biases in the
event reweighting procedure used in the fitting process and background
levels.  We obtain a coupling constant of
$|g_{CR}|=(0.100 \pm0.018_{STAT} \pm0.013_{SYS})$, corresponding to a
charge radius of $-\left\langle R^2 \right\rangle = 
(-0.047 \pm0.008_{STAT} \pm0.006_{SYS}) fm^2$.

\section{$\Kl\to\ppee$}

The decay $\Kl\to\ppee$ has no BE contributions to the amplitude and the
DE is suppressed because Bose statistics and gauge invariance force the
$\pi\pi$ system to have $l=2$.  As a result,
this process is dominated by the charge radius process studied above.
Existing predictions \cite{cpt} are in the range
$(0.8 \sim 2.0) \times 10^{-10}$.  The background
is $\Kl\to\pi^0\pi^0\pi^0$, where one of the $\pi^0$ decays results in
an $e^+e^-$ pair, possibly due to the interaction of a photon with
material in the detector.  One of the photons must go unseen, and so the
background is suppressed by requiring correctly reconstructed
$\pi^0$ masses and that the decay vertex found from the photons using a
$\pi^0$ mass constraint be consistent with the vertex found from the
$e^+e^-$ pair.  Using a Monte Carlo sample three times the size of the
data sample, we estimate the background to be
0.4\pma{0.4}{0.3} \ events.

\begin{figure}
\caption{$\Pt2$ {\it vs.} reconstructed mass in the data for the
$\ppee$ channel.}
\label{fig:valarie}\begin{center}
\epsfig{file=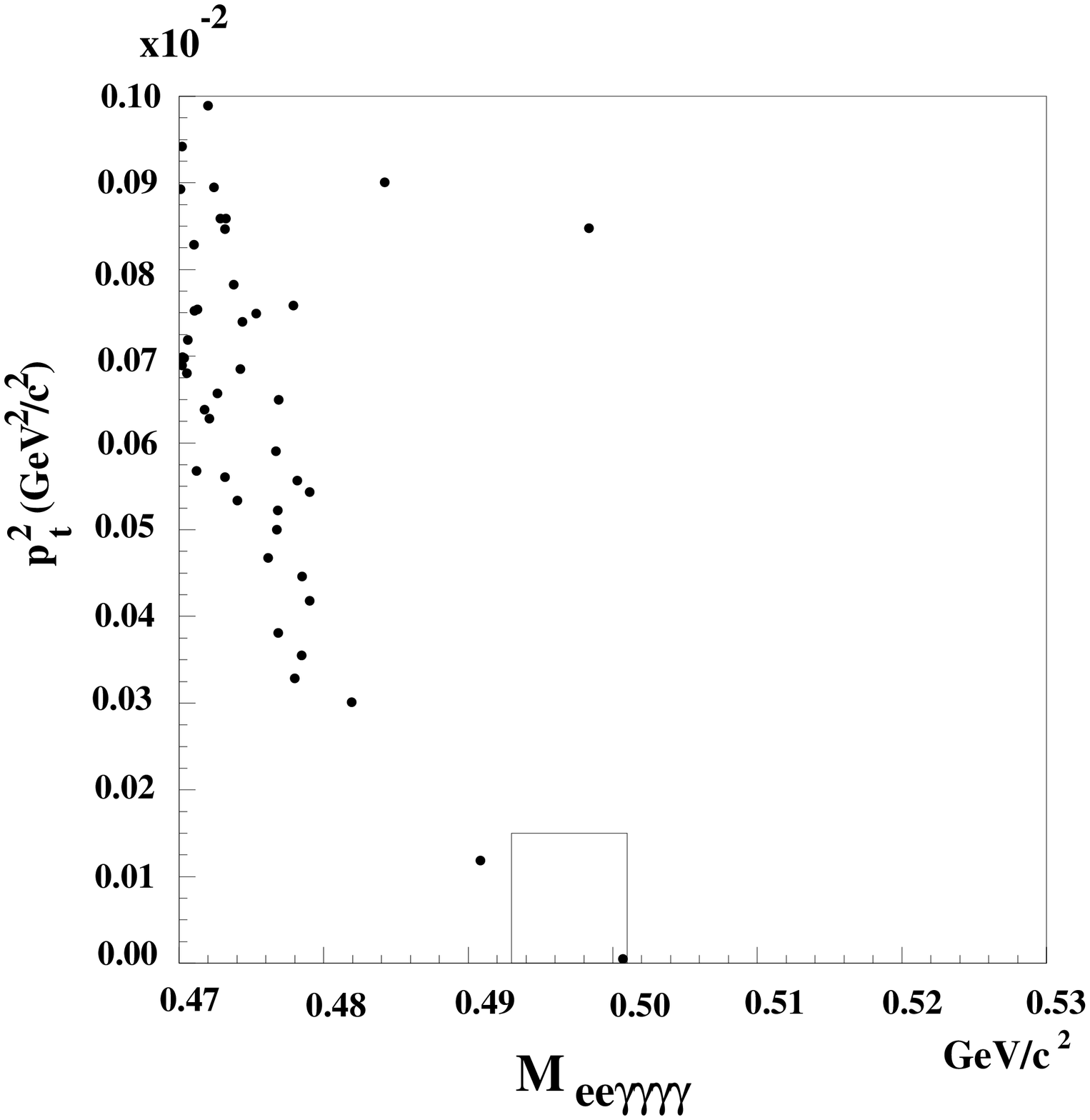,height=3.0in}
\end{center}\end{figure}

Figure~\ref{fig:valarie} shows the distribution of reconstructed
momentum transverse to the $\Kl$ flight direction squared ($\Pt2$)
versus the reconstructed mass in the data.  There is one event in the
signal region, and our preliminary 90\% C.L. limit, based on the 1997
data, is
\begin{equation}
Br(\Kl\to\ppee) < 5.4 \times 10^{-9},
\end{equation}
which is the first experimental result in this mode.  A short letter is
being prepared \cite{sasha}.

\section{Constraining $\row$}

The possibility of constraining or even measuring $\row$ with kaon
decays has been studied for quite some time.  The approach is to limit
the short-distance contributions to $\Kl\to\mu^+\mu^-$ by comparing the
measured branching ratio for this process with known and large
long-distance contributions which proceed through two photons.  To
understand these long-distance effect the decays $\Kl\to\mmee$,
$\Kl\to\eeee$, $\Kl\to e^+e^-\gamma$ and $\Kl\to\mmg$ are
of particular interest.  At this time, the best limits on $\row$ are
obtained using form factors from the $\Kl\to\mmg$ channel, but these
limits are much weaker than those existing from other global analyses.
A more leisurely discussion is available in \cite{myself} and the
references therein.

\section{Lepton Flavor Violation}

\begin{figure}
\caption{$\Pt2$ {\it vs.} reconstructed mass in the data for the
$\pi^0\mu^\pm e^\mp$ channel.}
\label{fig:angela}\begin{center}
\vspace{0.10in}
\epsfig{file=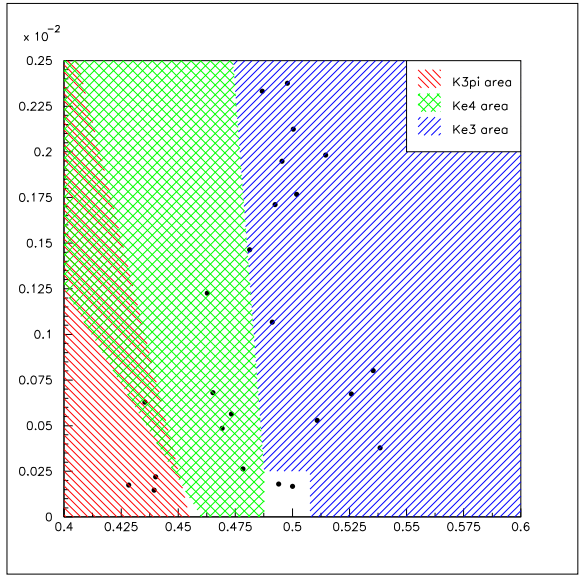,height=3.0in}
\end{center}\end{figure}

At KTeV, we can search for the lepton flavor violating process
$\Kl\to\pi^0\mu^\pm e^\mp$.  Figure~\ref{fig:angela} shows the
distribution of $\Pt2$ versus the reconstructed mass in the 1997 data.
The backgrounds are $\Kl\to\pi^+\pi^-\pi^0$,
$\Kl\to\pi^0\pi^\pm e^\mp\nu$, and $\Kl\to\pi^\pm e^\mp\nu$ with other
activity in the detector that appears as a $\pi^0$.  The regions where
these different backgrounds predominate are marked in
Fig. \ref{fig:angela}; we estimate a background level of $0.61\pm0.56$
events.  There are two events in the signal region, and our preliminary
90\% C.L. limit is
\begin{equation}
Br(\Kl\to\pi^0\mu^\pm e^\mp) < 4.4 \times 10^{-10}.
\end{equation}

\section{More Data!}

The data taken in 1999 will increase the KTeV data set by a factor of
about 2.3.  The kaon physics community in the U.S. is increasingly
focused on the $\pi\nu\overline\nu$ modes, which will permit placing
theoretically clean constraints upon $(\row,\ate)$, and possibly the
discovery of new physics if those constraints are incompatible with
other results, such as the recent B factory results.  Two
collaborations proposed to study these modes using
the Fermilab Main Injector; CKM proposed \cite{homepage} to measure the
$K^+\to\pi^+\nu\overline\nu$ mode and KaMI proposed to measure
$\Kl\to\pi^0\nu\overline\nu$.  After this conference, both programs
were extensively reviewed.  CKM won scientific approval from Fermilab,
but KaMI unfortunately did not.

\end{document}